\documentclass[12pt]{article}
\usepackage{amssymb}
\usepackage{amsmath}
\usepackage{graphicx}

\def\beq{\begin{eqnarray}}    
\def\eeq{\end{eqnarray}}      

\def\ln{\,\mbox{ln}\,}                  

\def\im{\textrm{i}}

\def\sfrac#1#2{{\textstyle\frac{#1}{#2}}}
\def\={\ =\ }

\def\vp{\varepsilon}

\begin{document}

\begin{titlepage}
\setcounter{page}{0}

\vskip 2.0cm

\begin{center}

{\LARGE\bf Notes on soft breaking of BRST symmetry in the
Batalin-Vilkovisky formalism }

\vspace{18mm}

{O.V.Radchenko$\,{}^{\dagger}$, \
A.A. Reshetnyak$\,{}^\ast$
}

\vspace{8mm}

\noindent ${}^\dagger${\em Tomsk State Pedagogical University,
Kievskaya St.\ 60, 634061 Tomsk, Russia }

\vspace{4mm}

$^\ast${\em
Institute of Strength Physics and Material Science, \\
Akademicheskii av. 2/4, 634021 Tomsk, Russia}

\vspace{18mm}

\begin{abstract}
We have proved the nilpotency of the operators which
describe the gauge dependence of the generating functionals of
Green's functions for the gauge theories with the soft breaking of
BRST symmetry in the Batalin-Vilkovisky formalism.
\end{abstract}

\end{center}

\vfill \noindent{\sl E-mails:} \
${}^\dagger$radchenko@tspu.edu.ru, $\,{}^\ast$reshet@ispms.tsc.ru\\


\end{titlepage}


\section{Introduction}
 In our paper we consider the problems related to the dependence
 of the Green's functions on the gauge in the soft breaking of BRST symmetry
 in the Batalin-Vilkovisky formalism  \cite{BV} proposed in our previous
 papers \cite{llr}, \cite{lrr}.

This breakdown in Yang-Mills theories is connected with a
restriction of the domain of integration in the functional integral
due to the Gribov horizon \cite{Gribov} and introducing of the
Gribov-Zwanziger action \cite{Zwanziger1},\cite{Zwanziger2}. Note, the
investigations for the theories above have been performed, as a rule, in the
Landau gauge only (see, e.g., \cite{sorella1} and references
therein).

At the same time, it is well known fact that the physical quantities
and, in particular, S-matrix, can be calculated in different gauges
in the framework of the Batalin-Vilkovisky of quantization of gauge
theories, but they must not depend on a choice of the gauge
condition.  Recently, in Refs. \cite{llr}, \cite{lrr}  a
generalization of  the definition of soft breaking of BRST symmetry
valid for general gauge theories in arbitrary gauges within
field-antifield formalism has been proposed.

The aim of the paper is the detailed elaboration of the properties
of operators used in Ref. \cite{lrr} for expressing the variations
of generating functional of Green's functions under  variation of
the gauge condition.

We will use the condensed DeWitt's notations \cite{DeWitt}.
Derivatives with respect to sources and antifields are taken from
the left, while those with respect to fields are taken from the
right. The Grassmann parity of any quantity A in case of its
homogeneity is denoted as $\varepsilon (A)$.

\section{Odd Operators in BV quantization scheme with soft Breaking of BRST symmetry}
Consider the configuration space parameterized by the fields
 $\Phi\ \equiv\{\Phi^A\}\=\{A^i,\ldots\}$ with $\varepsilon(\Phi^A)=\varepsilon_A$,
  where the dots indicate the full set of additional to 
  $A^i$ fields of this theory in the BV method in dependence on its reducibility stage.
  Then for each field  $\Phi^A$ of this total configuration space, one should introduce the corresponding antifield  ~$\Phi^*$
  with opposite Grassmann parities to that of the corresponding
field $\Phi^A$ $\Phi^*\ \equiv\ \{\Phi^*_A\} \= \{A^*_i,\ldots\}$
with $\vp(\Phi^*_A)=\vp_A{+}1$.

In Ref. \cite{lrr} it was shown that for the generating functional
of Green's functions  $Z(J,\Phi^*)$,
\beq \label{ZBV} Z(J,\Phi^*)\=\int\!D\Phi\ \exp
\Big\{\frac{\im}{\hbar} \big(S(\Phi,\Phi^*)+ J_A\Phi^A\big)\Big\}\ ,
\eeq
with action  $S(\Phi,\Phi^*)$, which is additive extension of the
non-degenerate gauge-fixing action
 $S_{ext}(\Phi,\Phi^*)$ by the  bosonic functional   $M(\Phi,\Phi^*)$,
the variation of $Z(J,\Phi^*)$ induced by variation of the gauge is
written in the form
 \beq\label{varZ1} \nonumber \delta Z(J,\Phi^*)
&=& \frac{\im}{\hbar}\Big[\Big(J_A+M_{A}
\big(\sfrac{\hbar}{\im}\sfrac{\delta}{\delta
J},\Phi^*\big)\Big)\left(\frac{\delta}{\delta\Phi^*_A}\,
 -\frac{\im}{\hbar}
M^{A*} \big(\sfrac{\hbar}{\im}\sfrac{\delta}{\delta
J},\Phi^*\big)\right)\,\delta\Psi
\big(\sfrac{\hbar}{\im}\sfrac{\delta}{\delta J}\big) +\\
&& + \delta M\big(\sfrac{\hbar}{\im}\sfrac{\delta}{\delta
J},\Phi^*\big)\Big]Z(J,\Phi^*).  \eeq At deriving (\ref{varZ1}), we
have  taken into account that functionals  $S_{ext}(\Phi,\Phi^*)$
and $(-M)$ satisfy the quantum master-equations of the BV method,
$J_A$ appears by the sources to the fields ~$\Phi^A$, \ $\vp(J_A) = \vp_A$
and the notations  \beq \nonumber
M_{A}\big(\sfrac{\hbar}{\im}\sfrac{\delta}{\delta
J},\Phi^*\big)\equiv \frac{\delta M(\Phi,\Phi^*)}{\delta
\Phi^A}\Big|_{\Phi\rightarrow \frac{\hbar}{\im}\frac{\delta}{\delta
J}}, \qquad M^{A*}\big(\sfrac{\hbar}{\im}\sfrac{\delta}{\delta
J},\Phi^*\big)\equiv \frac{\delta
M(\Phi,\Phi^*)}{\delta\Phi^*_A}\Big|_{\Phi\rightarrow
\frac{\hbar}{\im}\frac{\delta}{\delta J}} \eeq were introduced. Here
$M=M(\Phi,\Phi^*)$ plays the role of the functional, which describes
the soft breaking of BRST symmetry in  \cite{lrr}.

Let us introduce the odd operator  $\hat{q}$: 
\beq \hat{q} & = & \Big(J_A+M_{A}\Big)\Big(\frac{\delta
}{\delta\Phi^*_A}-
\frac{\im}{\hbar}M^{A*}\Big),\label{q} \eeq 
which contains non-vanishing terms
$M_{A}\big(\sfrac{\hbar}{\im}\sfrac{\delta}{\delta J},\Phi^*\big)$,
$ M^{A*}\big(\sfrac{\hbar}{\im}\sfrac{\delta}{\delta J},\Phi^*\big)$
that differs it from the analogous operator considered in
\cite{llr}. Then $\delta Z(J,\Phi^*)$ in (\ref{varZ1}) can be
written in the form \beq\label{varZ2} \nonumber \delta Z(J,\Phi^*) =
\frac{\im}{\hbar}\Big[\hat{q}\,\delta\Psi
\big(\sfrac{\hbar}{\im}\sfrac{\delta}{\delta J}\big) + \delta
M\big(\sfrac{\hbar}{\im}\sfrac{\delta}{\delta
J},\Phi^*\big)\Big]Z(J,\Phi^*). \eeq

Let us prove the  nilpotency of the operator  $\hat{q}$,
i.e. that, $\hat{q}^2 = 0$.

To do this, we will use for the shortness   the following
notations: \beq \nonumber
M_{A}=M_{A}\big(\sfrac{\hbar}{\im}\sfrac{\delta}{\delta
J},\Phi^*\big), \qquad
M^{A*}=M^{A*}\big(\sfrac{\hbar}{\im}\sfrac{\delta}{\delta
J},\Phi^*\big). \eeq 
The square of  $\hat{q}$ may be directly presented as a sum of four
operators
 \beq \label{q2}\hat{q}^2 & = &
\Big[\Big(J_A+M_{A}\Big)\Big(\frac{\delta
}{\delta\Phi^*_A}- \frac{\im}{\hbar}M^{A*}\Big)\Big]^2\ \equiv\ \sum_{i=1}^4 D_i=\nonumber \\
& = &  \Big(J_A+M_{A}\Big)\frac{\delta
}{\delta\Phi^*_A}\Big(J_B+M_{B}\Big)\frac{\delta }{\delta\Phi^*_B} \
-\  \frac{\im}{\hbar} \Big(J_A+M_{A}\Big)\frac{\delta
}{\delta\Phi^*_A}\;\Big(J_B+M_{B}\Big)M^{B*}- \\
&-&\  \frac{\im}{\hbar}\Big(J_A+M_{A}\Big)
M^{A*}\Big(J_B+M_{B}\Big)\frac{\delta }{\delta\Phi^*_B} +
\Big(\frac{\im}{\hbar}\Big)^2\Big(J_A+M_{A}\Big)M^{A*}\Big(J_B+M_{B}\Big)\;
M^{B*}\nonumber. \eeq Consider the first term in the decomposition
(\ref{q2}),
 \beq \label{D1in}D_1 & = &  \Big(J_A+M_{A}\Big)\frac{\delta M_{B}
}{\delta\Phi^*_A}\frac{\delta}{\Phi^*_B}+\Big(J_A+M_{A}\Big)\Big(J_B+M_{B}\Big)\frac{\delta
}{\delta\Phi^*_B}\frac{\delta }{\delta\Phi^*_A}(-1)^{\vp_A{+}1}.
\eeq In turn, the second summand in  $D_1$ (\ref{D1in}) has the form
 \beq \label{D1} (-1)^{\vp_A+1}
\big(J_AJ_B +M_{A}M_{B} + J_AM_{B}+ (-1)^{\vp_A\vp_B} J_BM_{A} +
\sfrac{\hbar}{\im}M_{AB}\big)\frac{\delta
}{\delta\Phi^*_B}\frac{\delta }{\delta\Phi^*_A}\ , \eeq 
where we have taken into account that  $[M_A,\, J_B]=\frac{\hbar}{\im}M_{AB}$, determined as,  
\beq M_{AB} \= \frac{\delta^2 M(\Phi,\Phi^*) }{\delta\Phi^A\
\delta\Phi^B}\Big|_{\Phi \to \frac{\hbar}{\im}\frac{\delta}{\delta
J}} \qquad\textrm{and}\qquad M_{AB}=(-1)^{\vp_A\vp_B}M_{BA}\ . \eeq

Note, the generalized symmetry properties of the expression in the
brackets in (\ref{D1}) and the second derivative with respect to
antifields are opposite under replacement of  indices
$A\leftrightarrow B$ . From this fact it follows that (\ref{D1}) is
equal to zero and  $D_1$ term is determined only by the first term.
Turning to the summand $D_2$ we see that after rearranging of the
derivatives with respect to antifields $D_2$ takes the form
 \beq \label{D2} D_2
& = & -\  \frac{\im}{\hbar} \Big(J_A+M_{A}\Big)\frac{\delta M_{B}
}{\delta\Phi^*_A}M^{B*} - \frac{\im}{\hbar}
\Big(J_A+M_{A}\Big)\Big(J_B+M_{B}\Big)\frac{\delta M^{B*}
}{\delta\Phi^*_A}\;(-1)^{(\vp_A+1)\vp_B}-\nonumber \\
&& -\ \frac{\im}{\hbar}
\Big(J_A+M_{A}\Big)\Big(J_B+M_{B}\Big)M^{B*}\frac{\delta
}{\delta\Phi^*_A}\;(-1)^{\vp_A+1}.\eeq Then, we can see  taking into
account of the generalized symmetry property
 \beq \frac{\delta M^{B*}
}{\delta\Phi^*_A}=\frac{\delta M^{A*} }{\delta\Phi^*_B}
(-1)^{(\vp_A+1)(\vp_B+1)},\eeq that  the second term in $D_2$
vanishes.

Next, for the term  $D_3$ we have after sequence of the
transformations
 \beq D_3 & = & - \frac{\im}{\hbar} \Big(J_A+M_{A}\Big)M^{A*}J_B\frac{\delta
}{\delta\Phi^*_B}- \frac{\im}{\hbar}
\Big(J_A+M_{A}\Big)M^{A*}M_{B}\frac{\delta }{\delta\Phi^*_B}
=\nonumber \\
& = & - \frac{\im}{\hbar}
\Big(J_A+M_{A}\Big)\Big(\frac{\hbar}{\im}M_{\;\;\;B}^{A*}+J_BM^{A*}(-1)^{\vp_B(\vp_A+1)}\Big)\frac{\delta
}{\delta\Phi^*_B} -\\ &&- \frac{\im}{\hbar}
\Big(J_A+M_{A}\Big)M^{A*}M_{B}\frac{\delta
}{\delta\Phi^*_B}=\nonumber \\ &=&
-\Big(J_A+M_{A}\Big)M_{\;\;\;B}^{A*}\frac{\delta }{\delta\Phi^*_B}-
\frac{\im}{\hbar}
\Big(J_A+M_{A}\Big)\Big(J_B+M_{B}\Big)M^{A*}\frac{\delta
}{\delta\Phi^*_B}(-1)^{\vp_B(\vp_A+1)},\nonumber \eeq where we have
used the notation  \beq M_{\;\;\;B}^{A*}\=\frac{\delta^2
M(\Phi,\Phi^*)}{\delta\Phi^*_A\ \delta\Phi^B} \Big|_{\Phi\rightarrow
\frac{\hbar}{\im}\frac{\delta}{\delta J}}. \eeq

At last, transforming  the fourth summand  $D_4$ in (\ref{q2}) as
follows,
 \beq D_4
& = &
\Big(\frac{\im}{\hbar}\Big)^2\Big(J_A+M_{A}\Big)\Big(\frac{\hbar}{\im}
M_{\;\;\;B}^{A*}+J_B M^{A*}(-1)^{\vp_B(\vp_A+1)}\Big)M^{B*}+\nonumber \\
&
&+\Big(\frac{\im}{\hbar}\Big)^2\Big(J_A+M_{A}\Big)M^{A*}M^{B*}M_{B}=\nonumber \\
& =
&\frac{\im}{\hbar}\Big(J_A+M_{A}\Big)M_{\;\;\;B}^{A*}M^{B*}+\Big(\frac{\im}{\hbar}\Big)^2J_B
J_A M^{A*}M^{B*}(-1)^{\vp_B} +\nonumber \\
& &+ \Big(\frac{\im}{\hbar}\Big)^2\Big(\frac{\hbar}{\im} M_{AB}+J_B
M_{A}(-1)^{\vp_A\vp_B}\Big)M^{A*}M^{B*}(-1)^{\vp_B(\vp_A+1)}+\nonumber\\
&
&+\Big(\frac{\im}{\hbar}\Big)^2\Big(J_A+M_{A}\Big)M^{A*}M^{B*}M_{B}\nonumber,\eeq
we have finally,
 \beq\label{D4}D_4& =
&\frac{\im}{\hbar}\Big(J_A+M_{A}\Big)M_{\;\;\;B}^{A*}M^{B*}
+\Big(\frac{\im}{\hbar}\Big)^2J_B J_A M^{A*}M^{B*}(-1)^{\vp_B} +
\nonumber \\& &+\frac{\im}{\hbar} M_{AB}
M^{A*}M^{B*}(-1)^{\vp_B(\vp_A+1)}+\Big(\frac{\im}{\hbar}\Big)^2J_B
M_{A}M^{A*}M^{B*}(-1)^{\vp_B}-
\\ & &-\Big(\frac{\im}{\hbar}\Big)^2J_A
M_{B}M^{B*}M^{A*}(-1)^{\vp_A}+\Big(\frac{\im}{\hbar}\Big)^2
M_{A}M_{B}M^{A*}M^{B*}(-1)^{\vp_B(\vp_A+1)}. \nonumber\eeq

It is easy to see that the second, third and sixth terms in the last
expression identically vanish due to the generalized symmetry
properties  under changing of the indices  $A\leftrightarrow B$.
Indeed, the quantities $(J_B J_A)$, $M_{AB}$ are
generalized-symmetric ones, whereas  $(M^{A*}M^{B*}(-1)^{\vp_B})$ is
generalized-antisymmetric, and $(M_{A}M^{A*})^2 \equiv 0$. Next, the
sum of the fourth and fifth terms is equal to zero.  Therefore $D_4$
is reduced to the first term in  (\ref{D4}).

In view of the  derivations above, we have the following final
representations for $D_1$, $D_2$, $D_3$, $D_4$, \beq \label{D11} D_1
& = & \Big(J_A+M_{A}\Big)\frac{\delta M_{B}
}{\delta\Phi^*_A}\frac{\delta }{\delta\Phi^*_B},\\ \label{D12} D_2 &
= & -\ \frac{\im}{\hbar} \Big(J_A+M_{A}\Big)\frac{\delta M_{B}
}{\delta\Phi^*_A}M^{B*} - \frac{\im}{\hbar}
\Big(J_A+M_{A}\Big)\Big(J_B+M_{B}\Big)M^{B*}\frac{\delta
}{\delta\Phi^*_A}\;(-1)^{\vp_A+1},\\ \label{D13} D_3 & = &
-\Big(J_A+M_{A}\Big)M_{\;\;\;B}^{A*}\frac{\delta }{\delta\Phi^*_B}-
\frac{\im}{\hbar}
\Big(J_A+M_{A}\Big)\Big(J_B+M_{B}\Big)M^{A*}\frac{\delta
}{\delta\Phi^*_B}(-1)^{\vp_B(\vp_A+1)},\\ \label{D14} D_4 & =
&\frac{\im}{\hbar}\Big(J_A+M_{A}\Big)M_{\;\;\;B}^{A*}M^{B*}.\eeq
From the Eqs. (\ref{D11})-(\ref{D14}) we immediately obtain, first,
the sum of the operator  $D_4$ and the first term in $D_2$ is equal
to zero, second, the sum of the operator  $D_1$ and the first term
in  $D_3$ vanishes, third, the sum of the second terms in both
operators  $D_2$ and $D_3$ is equal to zero.

Thus, our statement, that  $\hat{q}^2=0$ is completely proved.

As the consequence, we have simultaneously proved the nilpotency of
the operator $\hat{Q}$, which is unitarily related to  $\hat{q}$
\beq \label{Q} \hat{Q} =
\exp{-\frac{\im}{\hbar}W}\hat{q}\exp{\frac{\im}{\hbar}W} =
\Big(J_A+M_{A}\Big(\frac{\delta W}{\delta
J}+\frac{\hbar}{i}\frac{\delta}{\delta
J},\Phi^*\Big)\Big)\frac{\delta }{\delta\Phi^*_A}.\eeq

Remind \cite{lrr}, the operator $\hat{Q}$ the dependence of the
generating functional of connected Green's functions $W(J,\Phi^*)=
\frac{\hbar}{\imath}\ln Z (J,\Phi^*)$ on the variation of the  gauge
in the representation
 \beq
 \delta
W(J,\Phi^*)&=& {\hat Q} \delta\Psi
 \Big(\frac{\delta W}{\delta J}+\frac{\hbar}{i}\frac{\delta}{\delta
 J},\Phi^*\Big) +
\delta M\Big(\frac{\delta W}{\delta
J}+\frac{\hbar}{i}\frac{\delta}{\delta J},\Phi^*\Big).
\label{varW}  %
\eeq At last, for the generating functional of vertex Green's
functions (effective action), which is obtained from $W(J,\Phi^*)$
by means of the Legendre transformation with respect to sources
 $J_A$, $\bigl(\Gamma
(\Phi,\;\Phi^*) = W(J,\Phi^*) - J_A\Phi^A\bigr)$ with average fields
 $\Phi^A = \frac{\delta W}{\delta J_A}$, the local (for $M=0$)
 representation for the odd and nilpotent operator $\hat{s}_q$ being equal to $\hat{Q}$,
 but acting on the space of average fields and antifields will be valid as well,
  \beq
\delta\Gamma (\Phi,\;\Phi^*) &=&  {\hat s}_q  <{\delta\Psi}>
+<\delta M>. \label{varGamma}
\eeq The angle brackets in the expressions  (\ref{varGamma}) denote
the averaging of the quantities and operators with respect to the
functional $\Gamma (\Phi,\;\Phi^*)$, considered in details in
\cite{llr}, \cite{lrr}, whereas the operator $\hat{s}_q$ itself has
the form, presented in fact in  \cite{lrr},with using the left
derivative with respect to  $\Phi^A$,
\beq {\hat s}_q & =& -\Bigl(\frac {\delta \Gamma}{\delta \Phi^{A}}
- {\hat M}_{A}\Bigr) \frac {\delta}{\delta \Phi^{*}_{A}}-
(-1)^{\vp_A}\Bigl(\frac {\delta \Gamma}{\delta \Phi_{A}^*} - {\hat
M}^{A^*}\Bigr)
\frac {\delta_l}{\delta \Phi^{A}} \nonumber \\
&&\quad -\ \frac{\im}{\hbar}\Big[{\widehat M}_A \frac
{\delta\Gamma}{\delta\Phi^*_A}+\frac{\delta\Gamma}{\delta
\Phi^A}{\widehat M}^{A*} -{\widehat M}_A{\widehat M}^{A*}  ,\
\Phi^B\Big\} \frac{\delta_{\it
l}}{\delta\Phi^B} \;. \label{hats} %
\eeq where the sign $[\ ,\ \}$ means for  supercommutator.

The nilpotency of the operators $\hat{q} $, $ \hat{Q} $, $
\hat{s}_q$ proved here repeats the properties of theirs analogs in
the Ref. \cite{llr}, but in case of more general regularization than
dimensional one and takes fundamental character reflecting the
presence of the BRST symmetry in the theory but with its soft
breaking.

\subsection*{Acknowledgement}
 The authors are thankful to P.M. Lavrov  for useful
discussions.
The work of O.V.Radchenko is supported by the  RFBR grant
12-02-31820. The work of A.A.Reshetnyak was supported  by
 the RFBR grant, project  Nr. 12-02-00121
and by LRSS grant Nr.224.2012.2.
\par\bigskip
\par
\end{document}